\documentclass[epjCONF]{svjour}
\usepackage{graphicx}
\usepackage[varg]{txfonts} % Times fonts
\usepackage[latin1]{inputenc}
\session-title{%
19$^{\textnormal{\footnotesize th}}$ International %
IUPAP Conference on Few-Body Problems in Physics%
}
\begin{document}
\title{%
%Insert your title here: 
Pionic Deuterium
}%
\author{%
% add email of the responsible author:
        Th. Strauch\inst{1},
        F.~D.~Amaro\inst{2}, 
        D.~F.~Anagnostopoulos\inst{3},
        P.~B\"uhler\inst{4}, 
        D.~S.~Covita\inst{2}\thanks{\emph{present address:} 
        Dept. of Physics, Aveiro University, P-3810 Aveiro, Portugal}, 
        H.~Gorke\inst{5}, 
        D. Gotta\inst{1}\thanks{\emph{corresponding author:} d.gotta@fz-juelich.de}, 
        A.~Gruber\inst{4}, 
        A.~Hirtl\inst{4}\thanks{\emph{present address:} 
        Universit\"atsklinik f\"ur Nuklearmedizin, Medizinische Universit\"at Wien, 1090 Vienna, Austria}, 
        P. Indelicato\inst{6},
        E.-O.~Le Bigot\inst{6},
        M. Nekipelov\inst{1},
        J.~M.~F.~dos Santos\inst{2},
        Ph.~Schmid\inst{4},
        S.~Schlesser\inst{6},
        L.~M.~Simons\inst{7},
        M.~Trassinelli\inst{6}\thanks{\emph{present address:} 
        Inst. des NanoSciences de Paris, CNRS UMR7588 and UMPC-Paris 6, F-75015 Paris, France}
        J.~F.~C.~A.~Veloso\inst{8},
   \and J.~Zmeskal\inst{4}
}
\institute{%
           Institut f\"ur Kernphysik, Forschungszentrum J\"ulich, D--52425 J\"ulich, Germany
           \and Dept. of Physics, Coimbra University, P-3000 Coimbra, Portugal
           \and Dept. of Materials Science and Engineering, University of Ioannina, GR-45110 Ioannina, Greece 
           \and Stefan Meyer Institut for Subatomic Physics, Austrian Academy of Sciences, A-1090 Vienna, Austria
           \and Zentralabteilung f\"{u}r Elektronik, Forschungszentrum J\"{u}lich GmbH, D-52425 J\"{u}lich, Germany
           \and Laboratoire Kastler Brossel, UPMC-Paris 6, ENS, CNRS; Case 74, 4 place Jussieu, F-75005 Paris, France 
           \and Laboratory for Particle Physics, Paul Scherrer Institut, CH--5232 Villigen, Switzerland
           \and Dept. of Physics, Aveiro University, P-3810 Aveiro, Portugal
}
\abstract{
The strong interaction shift $\epsilon^{\pi D}_{1s}$ and broadening $\Gamma^{\pi D}_{1s}$ in pionic deuterium 
have been remeasured with high statistics by means of the $\pi$D$(3p-1s)$ X-ray transition using the 
cyclotron trap and a high-resolution crystal spectrometer.  
Preliminary results are $\epsilon^{\pi D}_{1s}=(-2325\pm 31)$\,meV (repulsive) and 
$\Gamma^{\pi D}_{1s}(1171{+\,23\atop-\,49})$\,meV which yields for the $\pi$D scattering length 
$a_{\pi D}=[-(24.8\pm 0.3) +\,i\,(6.3{+\,0.1\atop -0.3})]\cdot 10^{-3}\,m^{-1}_{\pi}$. From the imaginary 
part the threshold parameter for pion production is determined to be $\alpha=(252{+\,5\atop -11})\,\mu$b. 
} %end of abstract
\maketitle
%
% 
%----- Beginning of MAIN TEXT  --------------------------------------- 
% 
\section{Introduction}

The deuterium experiment is embedded in a series of measurements\,\cite{PSI98,Got04,Got08} aiming at a 
precision determination of the pion-nucleon scattering length comprising
\begin{itemize}
\item pionic hydrogen ($\pi$H),
\item pionic deuterium ($\pi$D),
\item muonic hydrogen ($\mu$H), and
\item the characterisation of the Bragg crystal spectrometer using narrow X-ray lines from few-electron atoms 
produced by means of an electron-cyclotron resonance trap (ECRIT). 
\end{itemize}
In this contribution, first results of the $\pi$D measurement are presented.

\section{Strong-interaction effects}

The complex pion-nucleus scattering length $a_{\pi A}$ of a pionic atom formed with a nucleus $A(Z,N)$ is related 
in leading order to the 1s-state shift $\epsilon_{1s}$ and width $\Gamma_{1s}$ by the Deser 
formula\,\cite{Des54} 
\begin{equation}
\epsilon_{1s}-i\,\frac{\Gamma_{1s}}{2}= -\frac{2\alpha^{3}\mu^{2}c^{4}}{\hbar c}\,a_{\pi A} + ...
\end{equation}
Ellipsis stand for corrections to the Deser formula because $a_{\pi D}$ is determined from a Coulomb 
bound state\,\cite{Tru61}. 

In $\pi$H, the ground-state shift and broadening may be related to the $\pi N$ isoscalar and isovector scattering 
lengths $a^+$ and $a^-$ by 
\begin{eqnarray}
\epsilon^{\pi H}_{1s}&\propto & a_{\pi^{-}p\to\pi^{-}p}\hspace{4mm}= a^{+}+a^{-} ~~+~~...\nonumber\\
\Gamma^{\pi H}_{1s}  &\propto & (a_{\pi^{-}p\to\pi^{0}n})^2        = \,2(a^{-})^2\,~~~+~~...\,.
\end{eqnarray} 
In this case, $\Gamma \propto a^2$ because the hadronic broadening is due to a scattering process -- the charge-exchange 
reaction $\pi^-p\rightarrow \pi^0n$\,\cite{Ras82}.
The corrections here have been calculated in various approaches and amount up to a few per cent of the leading 
contribution and include to some extent already electromagnetic and strong isospin-breaking 
terms\,\cite{Sig96b,Lyu00,Gas02,Zem03,Oad06,Gas08}. 

In the case of $\pi$D, the real part $\Re\,a_{\pi D}$ of the scattering length, being proportional to the hadronic 
shift $\epsilon^{\pi D}_{1s}$,  maybe written by regarding the deuteron as a free proton and neutron 
as leading order and then applying additional corrections. One may write  
\begin{eqnarray}
\Re\,a_{\pi D}&=&a_{\pi^{-}p\to\pi^{-}p}+a_{\pi^{-}n\to\pi^{-}n}~~+~~...\nonumber\\
              &=&\hspace{12mm}2a^{+}\hspace{11mm}~~+~~...\,
\end{eqnarray} 
where ellipsis stand for the higher order terms from multiple scattering (depending also on $a^+$ and $a^-$), 
absorptive and possibly electromagnetic and strong isospin-breaking corrections. The second order correction is 
dominated by $(a^{-})^2$ being comparable in magnitude to the leading order term \cite{ThLa80,Eri88} because 
of the smallness of $a^+$ as required by chiral symmetry \cite{Wei66,Tom66}. 
The nuclear structure is taken into account by folding with the deuteron wave function. 

The pionic deuterium shift provides a constraint on the $\pi N$ scattering lengths $a^+$ and $a^-$. Its quality decisively 
depends on the precision of the data triple ($\epsilon^{\pi D}_{1s}$,$\epsilon^{\pi H}_{1s}$,$\Gamma^{\pi H}_{1s}$), 
for which a first set of precision data has been obtained in the last decade \cite{Sig96,Cha9597,Hau98,Sch01}. 
Vice versa, starting from the elementary pion-nucleon processes and involving properly nuclear structure, 
multiple scattering, and absorptive phenomena, the hadronic s-level shift of the $\pi$D system must be 
calculable unambiguously \cite{Bar97,Bea98,Del01,Eri02,Del03,Bea03,Bur03,Doe04,Mei05,Mei06,Val06}. 

The forthcoming results for $\pi$H from this measurement series will improve the constraint\,\cite{Gas08}, in 
particular by a better determination of the hadronic broadening in $\pi$H\,\cite{Got08}. Finally, the accuracy 
for $a^+$ will be determined mainly by the poorly known low-energy constant $f_1$ \cite{Gas08}.

Secondly, $\pi$D gives access to pion absorption and production $NN\leftrightarrow NN\pi$ at threshold \cite{Bru51}. 
The hadronic broadening $\Gamma^{\pi D}_{1s}$ mainly is due to true absorption  ($\pi^-d\rightarrow nn$) being the 
inverse and charge symmetric reaction of pion production  ($pp\rightarrow \pi^+d$) -- in contrast to $\pi$H where 
the width is exclusively due to charge-exchange and radiative capture. 

The relative strength of true absorption to radiative capture ($\pi^-d\rightarrow nn\gamma $) was measured to 
$S=\frac{nn}{nn\gamma}=2.83\pm\,0.04$\,\cite{Hig81}, and the branching ratio of internal pair conversion
($\pi^-d\rightarrow nne^+e^-$) and charge exchange ($\pi^-d\rightarrow nn\pi^0$) was found to be 0.7\%\,\cite{Jos60} 
and $1.45\pm 0.19\cdot 10^{-4}$\,\cite{Don77}, respectively. Hence, the relative strength of the true absorption 
channel to all other processes is obtained to $S'=nn/(nn\gamma+nne^+e^-+nn\pi^0)=2.76\pm\,0.04$, i.\,e., about 
2/3 of the hadronic width and with that of the imaginary part $\Im\,a_{\pi D}$ is related to the process 
$\pi^-d\rightarrow nn$, which can be linked with pion production. 

Pion production at low energies is usually parametrised by \cite{Ros54}
\begin{equation}
\sigma_{pp\rightarrow~\pi^{+}d}=\alpha C^{2}_{0}\eta + \beta C^{2}_{1}\eta^{3} + ...\label{eq:pid_pp}
\end{equation}
where $\eta =p^*_{\pi}/M_{\pi}$ is the reduced momentum of the pion in the $\pi d$ rest frame. 
For $\eta\rightarrow 0$ higher partial waves ($\beta,\,...$) vanish and only the threshold parameter 
$\alpha$ contributes representing pure s-wave production. The correction factors $C_{i}$ take into account 
the Coulomb interaction. $C^{2}_{0}$ is of the order of 30\%\,\cite{Rei69,Mac06} and an important source of 
uncertainty in the determination of $\alpha$ from cross-section data. 

Another approach to determine $\alpha$ is to exploit the $\pi$D ground state broadening where uncertainties 
stemming from Coulomb correction factors and normalisation of cross sections are avoided. To derive the relation 
between $\alpha$ and $\Im a_{\pi D}$ as obtained from pionic deuterium, purely hadronic (non-experimental) cross 
sections $\tilde{\sigma}$ are introduced to circumvent the problem of diverging Coulomb cross section at threshold. 
Detailed balance relates pion production and absorption by 
\begin{eqnarray}
\tilde{\sigma}_{\pi^{+}d\rightarrow\,pp}& = &\frac{2}{3}\cdot \left(\frac{p^{*}_{p}}{p^{*}_{\pi}}\right)^2\cdot \,\tilde{\sigma}_{pp\rightarrow~\pi^{+}d}
\end{eqnarray}
with $p^{*}_{p}$ and $p^{*}_{\pi}$ being final state center-of-mass (CMS) momenta\,\cite{Bru51}. 
Neglecting Coulomb and isospin breaking corrections, which are assumed to be at most a few per cent\,\cite{Bar09}, 
charge symmetry requires equal strength for the transitions $\pi^{-}d~\rightarrow~nn$ and $\pi^{+}d~\rightarrow~pp$. 

Combining optical theorem, charge invariance, detailed 
balance and inserting the parametrisation of the $pp\rightarrow \pi^+d$ cross section (\ref{eq:pid_pp}), the 
imaginary part of the $\pi^{-}d\rightarrow nn$ scattering length is related to the threshold 
parameter $\alpha$ by 
\begin{eqnarray}
\Im\,a_{\pi D}&=&(1+\frac{1}{S'})\cdot \Im\,a_{\pi^- d\rightarrow nn}\\
              &=&(1+\frac{1}{S'})\cdot\frac{p^*_{\pi}}{4\pi }\cdot \tilde{\sigma}_{\pi^-d\rightarrow nn}\nonumber\\ 
              &=&(1+\frac{1}{S'})\cdot\frac{1}{6\pi}\cdot \frac{(p^*_{p})^2}{m_{\pi}}\cdot\alpha\,.\label{eq:Ima_alpha} 
\end{eqnarray}
The factor $(1+1/S')$ corrects for the non true absorption channels.

For comparison of pionic-atom and pion-production data, the channels $pp\rightarrow~\pi^{+}d$ and 
$np\rightarrow~\pi^{0}d$ can be used because in the limit of charge independence 
$2\cdot\sigma_{np\rightarrow~\pi^{0}d}=\sigma_{pp\rightarrow~\pi^{+}d}$. Restricting to s waves, in both reactions 
the same transition of the nucleon pair $^{3}S_{1}(I=0)\rightarrow ^{3}P_{1}(I=1)$ occurs which corresponds also 
to pion absorption at rest on the deuteron's isospin 0 nucleon-nucleon pair.

\section{Atomic cascade}

\begin{figure}[b]
\resizebox{0.45\textwidth}{!}{\includegraphics{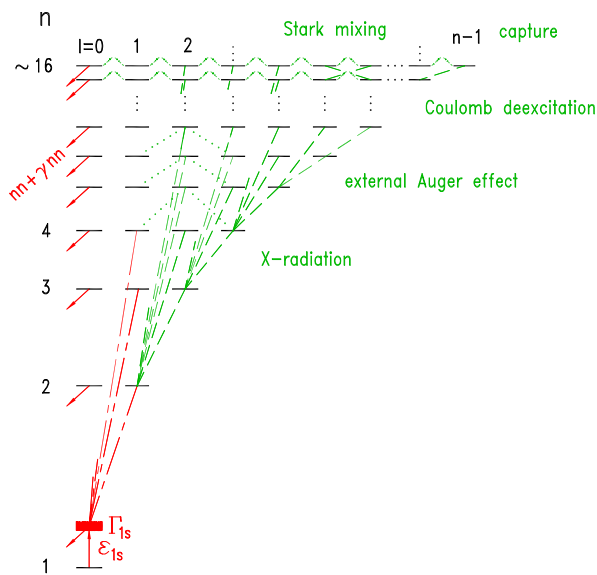}}
\caption{De-excitation cascade in pionic deuterium. The hadronic shift is defined to 
$\epsilon _{1s}\equiv E_{exp}-E_{QED}$, i.\,e., a negative sign as is the case in $\pi$D corresponds 
to a  repulsive interaction.}
%\vspace{4cm}
\label{figure:piD_cascade}
\end{figure}

After pion capture in hydrogen isotopes, a quantum cascade starts from main quantum numbers at about $n\approx 16$. The upper and medium part 
of the de-excitation cascade is dominated by collisional processes (Stark mixing, external Auger effect, Coulomb 
de-excitation). In the lower part X-radiation becomes more and more important (Fig.\,\ref{figure:piD_cascade}) 
\cite{Bor80,Har90,JeMa02}. Besides radiative de-excitation by X-ray emission all cascade processes depend on the 
environment, e.\,g., the density of the target gas. 

Stark mixing essentially determines the X-ray yields in exotic hydrogen. Because these systems are electrically 
neutral and small on the atomic scale, they penetrate surrounding atoms and, thus, experience a  
strong Coulomb field. Non-vanishing matrix elements $\langle nlm\mid {\bf E}\mid nl'm'\rangle$ in the presence 
of the electric field mix atomic states of the same principle quantum number {\em n} according to the selection 
rules $\Delta l=\pm 1$ and $\Delta m=0$\,\cite{Bet57}. An induced s state, from where pions 
disapper by nuclear reactions, leads to a depletion of the cascade. The Stark mixing rate is proportional to 
the number of collisions during the exotic atom's life time and explains the strong decrease of the X-ray yields 
with increasing target density\,\cite{Bor80}.
 
In the case of Coulomb de-excitation, the energy release of the de-excitation step is converted into kinetic 
energy of the collision partners\,\cite{BF78}. For lower-lying transitions a significant energy gain occurs 
which leads to a Doppler broadening of subsequent X-ray transitions. Doppler broadening was directly observed 
first in the time-of-flight spectra of monoenergetic neutrons from the charge exchange reaction at rest 
$\pi^-p\rightarrow \pi^0n$\,\cite{Czi63,Bad01}.

A dedicated measurement of the $\mu$H$(3p-1s)$ transition has been performed within this series of 
experiments\,\cite{Cov09}. In $\mu$H, where no hadronic broadening occurs, only the Doppler broadening 
exclusively from acceleration effects during the atomic cascade contributes and was directly observed as 
a significantly increased line width. The data were used -- among others -- to confirm that the methods used 
to quantify the corrections to the line width in pionic hydrogen and deuterium are sufficiently well 
defined \cite{Got08}. Details may be found elsewhere \cite{Covth}.

The acceleration due to Coulomb de-excitation is counteracted by elastic and inelastic scattering. This leads 
to a rather complex velocity distribution with tails below the peaks caused by the  Coulomb transitions which are 
at 12, 20, 38, and 81\,eV for the $\Delta n=1$ transitions $(7-6)$, $(6-5)$, $(5-4)$, and $(4-3)$, respectively. 

\begin{figure}[t]
\resizebox{0.45\textwidth}{!}{\includegraphics{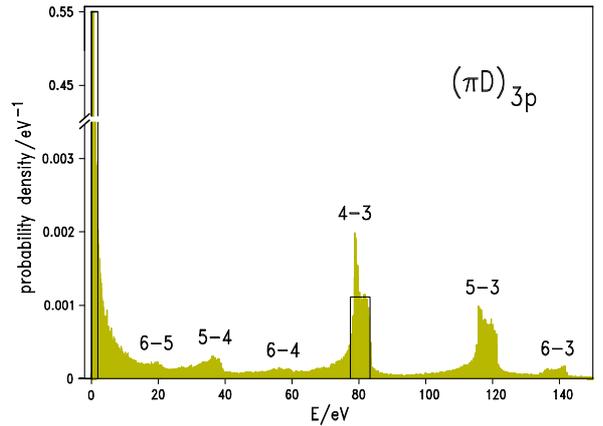}}
\caption{Prediction for the kinetic energy distribution of the $\pi$H atom at the instant of $(3p-1s)$ transition 
         scaled to the kinematics of the $\pi$D system for a density of 10\,bar equivalent. For the analysis, the 
         distribution was tentatively approximated by a box describing the low-energy component and a second one 
         for the high-energy component around 80\,eV. The low-energy component reaching 0.55 dominates the 
         distribution (note the broken vertical scale).}
%\vspace{4cm}
\label{figure:Tkin_ESCM}
\end{figure}

Cascade calculations have been extended to follow the development of the velocity during the 
de-ex\-citation cascade and, therefore, predict kinetic energy distributions at the time of X-ray emission from a 
specific level (extended standard cascade model ESCM \cite{JeMa02}). Calculations exist at present only for 
muonic and pionic hydrogen. Figure\,\ref{figure:Tkin_ESCM} shows such a prediction for $\pi$H where energies 
have been scaled to the $\pi$D case. It turned out in the analysis of the $\mu$H$(3p-1s)$ line 
shape that the first ESCM predictions could not describe the measured spectrum\,\cite{Cov09}. Therefore, 
a model independent approach was used to extract the relative strength of Doppler contributions directly from 
the measured line shape; a method applied successfully first in the neutron time-of-flight analysis\,\cite{Bad01}. 
Here, the kinetic energy distribution is modeled by boxes around the peaks generated by Coulomb transitions. 
One approach as used in the analysis of this experiment is included in Figure\,\ref{figure:Tkin_ESCM}. 

In the case of exotic hydrogen, Auger emission is only possible during the collision with other target atoms 
(external Auger effect). Auger transitions prefer de-excitation steps as small as possible given that the energy 
gain exceeds the binding energy of the electron. Consequently, Auger emission contributes mainly in the range 
$n\approx 6-10$ and  $\Delta n=1$ transitions are favored\,\cite{Bor80,JeMa02}. Due to the mass difference, 
Doppler broadening from recoil is negligibly small compared to the one caused by Coulomb transitions. 

It is known from muon-catalysed fusion experiments that during $\mu $H + H$_2$ collisions metastable hybrid molecules 
are formed like $[(pp\mu )p]ee$ \cite{Taq89,Jon99}. An analogue process is expected in $\pi $D + D$_2$ collisions,  
Such complex' are assumed to stabilise non radiatively by Auger emission. However, radiative decay out of molecular 
states has been discussed, and its probability is predicted to increase with nuclear mass \cite{Taq89,Jon99,Lin03,Kil04}. 
Possible X-ray transitions from molecular states would falsify the value for the hadronic shift determined from the 
measured X-ray energy. As molecular formation is collision induced, the fraction of molecules formed and the 
corresponding X-ray rate should depend on the target density. Therefore, a possiblity to identify such radiative 
contributions is a measurement of the X-ray energy at three different densities (3.3, 10, and 17.5\,bar equivalent 
density in this experiment).

\section{Experiment}

\begin{figure}[t]
\resizebox{0.45\textwidth}{!}{\includegraphics[angle=-90]{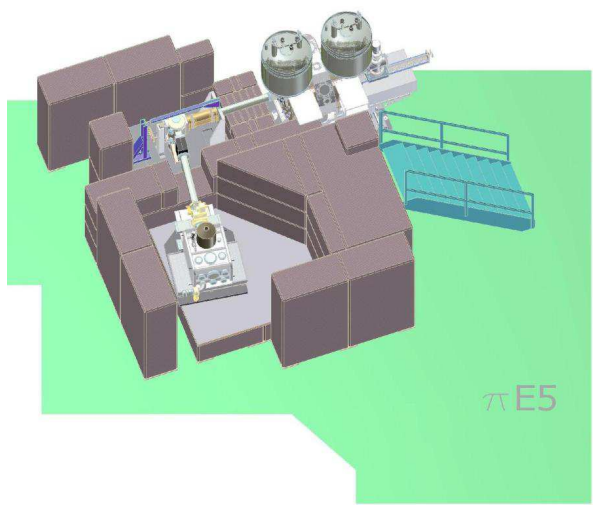}}
\caption{Setup of the $\pi$D experiment in the $\pi$E5 area at the Paul Scherrer Insitut (PSI). The roof 
         of the concrete shielding is omitted tp show the vacuum system connecting the cyclotron trap (upper 
         right), crystal chamber (upper left) and the cryostat of the X-ray detector (bottom left).}
%\vspace{4cm}
\label{figure:setup_piD}
\end{figure}

The experiment was performed at the $\pi$E5 channel of the proton accelerator at the Paul Scherrer Institut (PSI), 
Villigen, Switzerland, which provides a 
low-energy pion beam with intensities of up to a few $10^{8}$/s (Fig.\,\ref{figure:setup_piD}). Pions of 112 MeV/c 
were injected into the cyclotron trap II\,\cite{PSI98} and decelerated using a set of degraders optimized to 
the number of pion stops in a cylindrical cryogenic target cell of 22\,cm length and 5\,cm in diameter in the 
center of the trap. The cell was filled with deuterium gas cooled by means of a cold finger. About 
0.5\% of the incoming pions per bar equivalent pressure are stopped in $D_2$ gas. 
X-radiation could exit the target cell axially through a 5\,$\mu$m thick mylar$^{\textregistered}$ window. 
The window foil is supported by horizontal aluminum bars.

The Johann-type Bragg spectrometer was equipped with a 
spherically bent Si crystal cut parallel to the (111) plane having a radius of curvature of $R=2982.2\pm 0.3$\,mm. 
The reflecting area of the crystal having 10\,cm in diameter was restricted by a circular aperture to 95\,mm in 
diameter to avoid edge effects and to 60\,mm horizontally to keep the Johann broadening small\,\cite{Egg65,Zsc82}. 

Such a spectrometer is able to measure simultaneously an energy interval according to the width of the X-ray 
\linebreak source when using a correspondingly extended X-ray detector. Being a pixel device, CCDs are ideally 
suited for X-rays in the few keV range because they combine an intrinsic position resolution with the good energy 
resolution of semiconductor detectors. In addition, the granularity of the CCDs allows for efficient background 
rejection by means of pattern recognition (cluster analysis). Photo electrons from few-keV X-ray conversion are 
stopped within a few micrometer and, therefore, deposit charge in one or two pixels with a common boundary. 
Beam induced background, mainly high energy photons from neutrons produced in nuclear pion absorption 
and captured by surrounding nuclei, leads to larger structures. Together with the massive concrete shielding 
(Fig. \ref{figure:setup_piD}) such events are efficiently suppressed (Fig.\,\ref{figure:ADC_piD}). 

In this experiment, a $3\times 2$ array of charge-coupled devices (CCDs) was used covering in total 72\,mm in 
height and 48\,mm in width\,\cite{Nel02}. Monte Carlo studies show that about 2/3 of the intensity of the reflection 
is covered by the height of the array. The detector surface was oriented perpendicular to the direction 
crystal -- detector. 

The CCDs' depletion depth of about 30\,$\mu$m yields an optimum quantum efficiency in the 3-4\,keV range. A pixel 
size of 40\,$\mu$m provides a two-dimensional position resolution sufficient to measure precisely the shape of the 
diffraction image. The relative orientation of the individual CCDs as well as the exact pixel size was obtained by 
means of an optical measurement using a nanometric grid\,\cite{Ind06}. The CCDs are operated at a temperature of 
$-100^{\circ}$C. 

X-ray energies of the ground-state transitions in muonic and pionic hydrogen isotopes are in the few keV range, 
where no narrow $\gamma$ lines of sufficient intensity are available to be used to determine precisely 
the  function of a low-energy crystal spectrometer. However, from highly stripped atoms produced in an 
electron-cyclotron resonance trap (ECRIT) narrow X-ray lines are emitted at high rate.

The spectrometer response was measured using the narrow $M1$ X-ray line from helium-like argon of 3.104\,keV 
as described in\,\cite{Ana05,Tra07} yielding a resolution of 436$\,\pm$\,3\,meV (FWHM) when scaled to the 
energy of the $\pi$D$(3p-1s)$ transition of 3.075\,keV. This value is close to the intrinsic resolution of 
403\,$\pm$\,3\,meV given by the rocking curve width as calculated for an ideal flat crystal with the code 
XOP\,\cite{San98}. 

\begin{figure}[b]
\resizebox{0.45\textwidth}{!}{\includegraphics{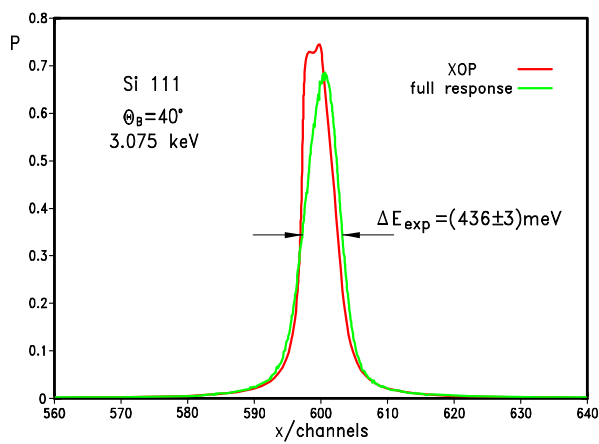}}
\caption{Transformation of the intrinsic resolution of the Si (111) reflection from a flat crystal by folding 
         with the experiment geometry and an additional Gaussian describing an imperfect mounting.}
%\vspace{4cm}
\label{figure:response}
\end{figure}

The response is constructed from the intrinsic properties of the Si (111) material as calculated from the 
dynamical diffraction theory with the code XOP\,\cite{San98}. The rocking curve is then convoluted with the 
experiment geometry by means of a Monte-Carlo ray-tracing code (Fig.\,\ref{figure:response}). Deviations from 
this ideal response are introduced by convoluting with an additional Gaussian contribution which was 
determined from the measurement of the $M1$ transition in helium-like argon.

In the Johann set-up the energy calibration must be provided by a reference line of known energy. In  order 
to avoid any mechanical changes, only X-rays having about the same Bragg angles are suitable for precision 
measurements. The best angular matching for the $\pi$D$(3p-1s)$ measurement was found for the gallium K$\alpha$ 
fluorescence lines. For energy calibration, the Ga K$\alpha_2$ transition (9224.484 $\pm$ 0.027\,eV\,\cite{Des03}) 
was chosen because of the smaller experimental uncertainty. The energy of the $\pi$D$(3p-1s)$ line is 
obtained by the angular distance to the Ga reference line to be calculated from the position difference of the 
reflections on the detector and the distance crystal -- detector. 

The fluorescence target itself was made of a $25\times 20$\,mm$^2$ GaAs plate mounted in the rear part of the 
gas cell but outside the pion stop volume at 82\,mm off the center of the cyclotron trap and away from the crystal. 
The X-rays were excited by means of an X-ray tube mounted on a window of the cyclotron trap chamber below the 
gas cell. The value for the tabulated Ga K$\alpha$ energies\,\cite{Des03} was obtained from measurement also using the 
compound GaAs\,\cite{Moo}. Therefore, a possible chemical shift must not be considered. 

The distance center of the crystal to center of the cyclotron trap was 2100\,mm being about 10\% outside the Rowland 
circle fulfilling the focal condition $R\cdot \sin\Theta_{B}$. The advantage placing the X-ray source off the 
focal position is that one averages over non uniformities of the target. Both GaAs and D$_2$ target are large 
enough that no cuts in the tails of the reflection occur. The distance crystal -- detector, chosen close to  
the assumed $\pi$D focal length, was determined to y$_{CD}=1918.1\,\pm\,0.5$\,mm by a survey measurement.

Alternating measurements of the Ga fluorescence radiation and the $\pi$D line were performed at least once per 
day (Fig.\,\ref{figure:GaKa_piD10bar}). The Ga fluorescence X-rays  together with two inclinometers mounted 
at the crystal and detector chambers were used to monitor the stability of the line position. Details on 
the experimental setup maybe found elsewhere\,\cite{Str09,Str10}.

\section{Analysis}

\begin{figure}[t]
\resizebox{0.45\textwidth}{!}{\includegraphics{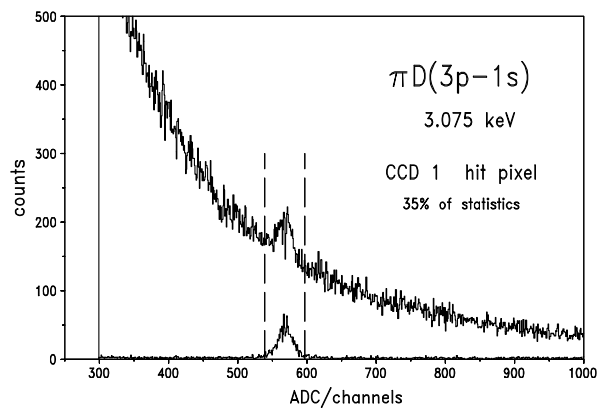}}
\caption{Background suppression achieved by the cluster analysis of the $\pi$D raw data.}
\label{figure:ADC_piD}
\end{figure}

The X-ray detector raw data are made available as the digitalised charge contents and a position index of the pixel. 
At first, the cluster analysis is performed. As expected, at 3.1\,keV only single ($\approx$\,75\%) or two pixel 
events ($\approx$\,25\%) contribute. Hot and defect pixels are masked by software. The cluster analysed charge (ADC) 
spectra of the CCDs show a pronounced peak originating from $\pi$D X-rays on a largely suppressed background 
(Fig.\,\ref{figure:ADC_piD}). For each CCD, an individual energy calibration was performed because of the different 
gain and noise behaviour. The energy resolution in terms of charge is determined by means of a Gaussian fit. 

Applying an energy cut in the ADC spectra, an additional background 
reduction is achieved. Several ADC cuts with a width of 1$\sigma$ to 4$\sigma$ of the respective Gaussian width were 
used to study the influence of peak-to-background on the result for the hadronic broadening. The minimum relative 
statistical error is achieved for a cut of 2.5$\sigma$ of the detector's (charge) resolution. With this energy cut, 
from fits to the position spectra (similar to the one shown in Figure \,\ref{figure:GaKa_piD10bar} -- bottom) count 
rates of 1448$\pm$49, 4010$\pm$74, and 4877$\pm$80 were obtained for the $\pi$D$(3p-1s)$ line at the equivalent 
densities of 3.3, 10, and 17.5\,bar, respectively. 

In the case of the Ga calibration line, compact clusters up to size 9 were accepted. Because of the larger 
penetration depth (5 and 105\,$\mu$m for 3.075 and 9.224\,keV in silicon\,\cite{Vei73}, respectively), X-rays 
also convert at the boundary of the depletion region where charge diffusion is already significant. As the 
calibration measurements were performed without pion beam, the number of background events is negligibly small 
and, hence, no suppression algorithm is necessary to clean the Ga spectra.

\begin{figure}[h]
\resizebox{0.47\textwidth}{!}{\includegraphics[angle=90]{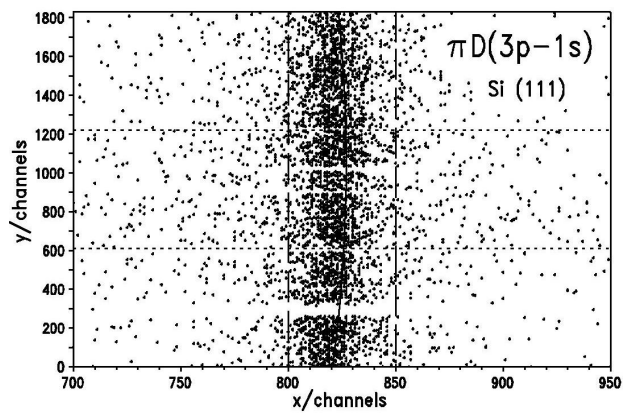}}
\caption{$\pi$D$(3p-1s)$ reflection after correction for curvature. The original curvature is indicated 
         by the parabola in the scatterplot of the already corrected data. One channel (pixel) in the 
         direction of dispersion ($x$) corresponds in first order reflection to 76.402$\pm$0.001\,meV.}
\label{figure:scat_piDline}
\end{figure}

The resulting hit pattern shows curved reflections on the CCD surface. The curvature is corrected for 
by means of a parabola fit (Fig.\,\ref{figure:scat_piDline}) before projecting onto the axis of dispersion, 
which is equivalent to an energy axis.

\subsection{Energy calibration}

The position of the Ga K$\alpha_2$ calibration line was determined applying a single Voigt profile in the 
fit. The same procedure was used for the tabulated values of the Ga K$\alpha$ X-ray energies\,\cite{Moo}. 

\begin{figure}[b]
%\begin{center}
\resizebox{0.45\textwidth}{!}{\includegraphics{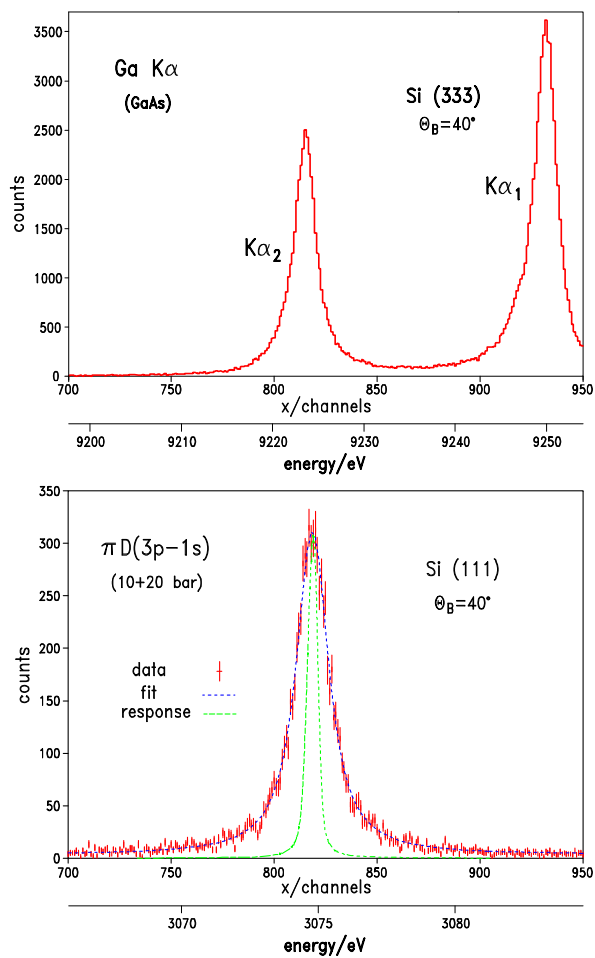}}
\caption{Ga K$\alpha_2$ calibration line and $\pi$D$(3p-1s)$ transition measured at an equivalent density of 10\,bar. 
         The narrow structure shown inside the $\pi$D line represents the resolution of the crystal spectrometer.}
%\end{center}
\label{figure:GaKa_piD10bar}
\end{figure}

The $\pi$D line was modeled both with a Voigt profile, where Doppler broadening and response function are 
approximated by a single Gaussian, and secondly, the true response as determined from the ECRIT data 
convoluted with the imaging properties and including the Doppler contributions from Coulomb de-excitation 
(Fig.\,\ref{figure:GaKa_piD10bar}). Both methods yield the same position value within a few hundreds of a pixel.

A significant correction to the Bragg angle is the index of refraction shift\,\cite{San98} because the $\pi$D and 
the Ga K$\alpha$ line is measured in first and third order, respectively. In addition, bending and penetration 
correction due to the substantially different energy must be considered. Mechanical shifts from temperature 
and vibrations are corrected by means of the inclinometer data.

In total, a systematic error of $\pm$\,8\,meV was deduced for the measurement of the $\pi$D$(3p-1s)$ energy. 
The combined statistical error obtained as weighted average from the data taken at the three different $D_2$ 
densities amounts to $\pm$\,11\,meV. The largest contribution to the error stems from the uncertainty of the 
energy of the Ga K$\alpha_2$ line of $\pm$\,27\,meV. For the transition energy, we find
\begin{eqnarray} 
E_{\pi D(3p-1s)}=3075.583\pm 0.030\,\mathrm{eV}.
\end{eqnarray}

\subsection{Line width}

To the line shape contribute the response of the spectrometer, the Doppler broadening from Coulomb transitions, 
and the natural line width of the $\pi$D$(3p-1s)$ transition. The natural line width is given by the width of the 
1s state of about 1\,eV because nuclear reactions from the 3p level are negligible and the radiative width is 
28\,$\mu$eV. 

The model free approach to identify possible Doppler contributions uses narrow boxes of e few eV width around the 
kinetic energies as suggested by the energy release of Coulomb transitions. Following the experience from the analysis 
of the $\mu$H$(3p-1s)$ line shape, one tries to identify consecutively individual contributions starting with only 
one contribution of the lowest possible energy, which corresponds to $\pi$D systems not accelerated or moderated 
down by collisions to energies of a few eV or below. It is sufficient to approximate the various kinetic energy 
components by box-like distributions as shown in the analysis of the $\mu$H experiment. 

\begin{figure}[t]
\resizebox{0.48\textwidth}{!}{\includegraphics{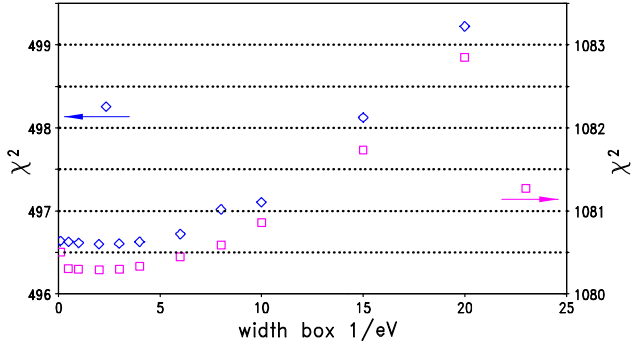}}
\caption{Search for evidence and width of the low-energy component in the kinetic energy distribution. 
         Diamonds (left $\chi^2$ scale) are due to the 10\,bar data. Squares (right scale) are from a 
         simultanuous fit to the two spectra taken at 10 and 17.5\,bar.}
\label{figure:searchbox1}
\end{figure}

A $\chi^2$ analysis, done by means of the MINUIT package\,\cite{Jam75}, shows that a low-energy contribution is 
mandatory. It was found that the kinetic energy must not exceed 8\,eV; a result found independently for taken 
at 10\,bar and the 17.5\,bar equivalent density (Fig.\,\ref{figure:searchbox1}). 
The result for the natural line width $\Gamma_{1s}$ turned out to be insensitive to the upper boundary of the 
low-energy box for values $\leq 8$\,eV. Therefore, the low-energy component was fixed to a width of 2\,eV in the 
further analysis. 

Noteworthy to mention that the ESCM prediction for the kinetic energy distribution for $\pi$H$(3p-1s)$ case 
scaled to $\pi$D kinematics (Fig.\,\ref{figure:Tkin_ESCM}) is unable to reproduce the $\pi$D$(3p-1s)$ line shape 
because of rather strong contributions of $(4-3)$ and $(5-3)$ Coulomb transitions. 
Searches for any higher energy contributions failed even when using the sum spectrum of the two measurements at 
10 and 17.5\,bar. This result is rather surprising because also the first analyses of the new pionic hydrogen 
data show that at least one high-energy component must be assumed to describe the line shape of the $\pi$H 
$(4-1)$, $(3-1)$, and $(2-1)$ transitions\,\cite{Got08,Hir08}. Though a significantly better description of 
the $\mu$H$(3p-1s)$ line shape could be achieved after a recalculation of collision cross 
sections\,\cite{Cov09,PP06,JPP07,PP07}, there is no explanation for the absense of any high-energy 
components in $\pi$D.

\begin{figure}[h]
\resizebox{0.45\textwidth}{!}{\includegraphics{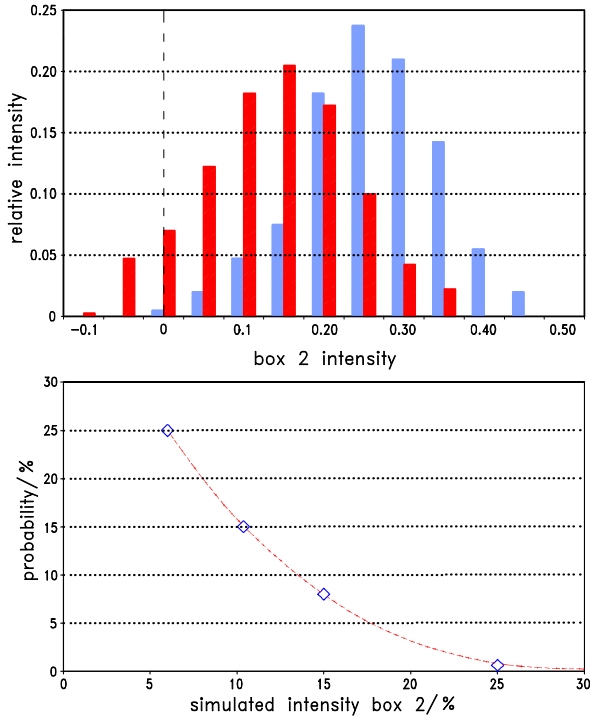}}
\caption{Top: distributions of relative intensity found from a fit with two boxes when a Doppler contribution 
         of 10\% and 25\% relative intensity, respectively, is assumed. The result is based on 400 Monte-Carlo 
         simulations each performed for the statistics of the sum of the 10 and 17.5\,bar data. 
         Bottom: probability to miss a Doppler contribution at 80\,eV as a function of its relative intensity.}
\label{figure:sensitivity}
\end{figure}

It has been studied in detail by Monte Carlo which magnitude of high-energy energy components maybe missed with 
the statistics achieved in this experiment. The probability to miss a high-energy component around 80\,eV (box\,2) 
corresponding to the $(4-3)$ Coulomb transition is displayed in Figure\,\ref{figure:sensitivity}. For each set of 
conditions 400 simulations were performed. It can be seen, that a high-energy contribution of 25\% or larger can 
hardly be missed. For 10\% relative intensity, the chance is about 15\% which corresponds to about 1$\sigma$ with 
respect to the full probability distribution allowing a certain variety for the shape of the box' extension 
(Fig.\,\ref{figure:sensitivity}).

\begin{figure}[t]
\resizebox{0.45\textwidth}{!}{\includegraphics{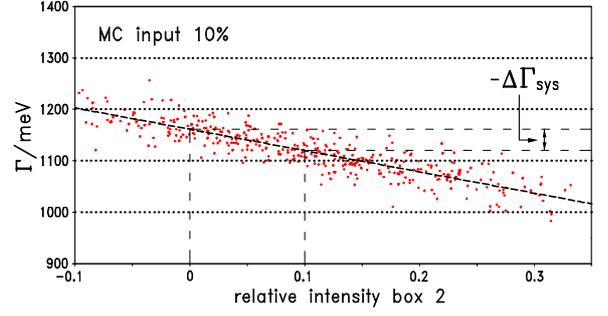}}
\caption{Distributions of relative intensity of a Doppler box around $T_{kin}=80$\,eV of 10\% relative intensity 
         versus the Lorentz width $\Gamma$ extracted from a set of 400 Monte-Carlo simulations performed for the 
         statistics of the sum of the 10 and 17.5\,bar data. The input value for $\Gamma=$1130\,eV.}
\label{figure:wolke}
\end{figure}

Assuming weight zero for box\,2 results in an upper limit for $\Gamma_{1s}$, and using the limit of sensitivity 
of 10\% for a second box yields a lower bound for $\Gamma_{1s}$ ($-\bigtriangleup \Gamma_{sys}$) according to the 
above-mentioned 1$\sigma$ criterion (Fig.\,\ref{figure:sensitivity}). 
The distribution of the results for the weight a second box (box 2 intensity) and the Lorentz width 
$\Gamma$ reflects the fluctuations due to the limited statistics (Fig.\,\ref{figure:wolke}). 

\section{Preliminary results}

For extracting the hadronic shift, the pure electromagnetic transition energy has been recalculated to be 
E$_{QED}$ =\linebreak
3077.9062 $\pm$ 0.0079\,eV. The error of this calculation is dominated by the uncertainty of the 
charged pion mass (0.0077\,eV)\,\cite{PDG08} where the accuracy of deuteron (0.0012\,eV) and  
pion radius (0.0010\,eV) contribute only marginal. The numerical accuracy of this calculation is assumed to 
be better than 1\,meV\,\cite{PI09}.

\begin{figure}[b]
\resizebox{0.48\textwidth}{!}{\includegraphics{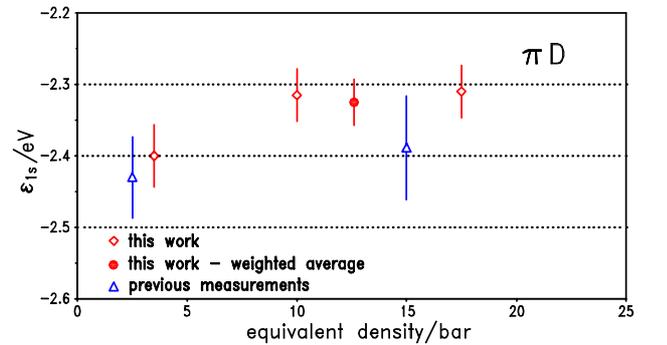}}
\caption{Experimental information on the hadronic shift in pionic deuterium. The error of the 
         individual measurements at 3.3, 10, and 17.5\,bar and, therefore, also of the weighted average 
         is dominated by the uncertainty of the Ga K$\alpha_2$ energy. The results from the two 
         previous experiments\,\cite{Cha9597,Hau98} are adjusted according to the new values for the 
         pure electromagnetic transition energies.}
\label{figure:shift}
\end{figure}

Combining measured and calculated QED transition energy, we obtain for the hadronic shift 
\begin{eqnarray}
\epsilon_{1s}=\,-\,(2323\,\pm\,31)\,\mathrm{meV}\label{eq:eps_final}. 
\end{eqnarray}
This result is compared to previous measurements in Figure\,\ref{figure:shift}. Note that the new result for the 
electromagnetic transition energy differs slightly from the values as used in the previous 
experiments\,\cite{Cha9597,Hau98}. 
 
\begin{figure}[t]
\resizebox{0.45\textwidth}{!}{\includegraphics{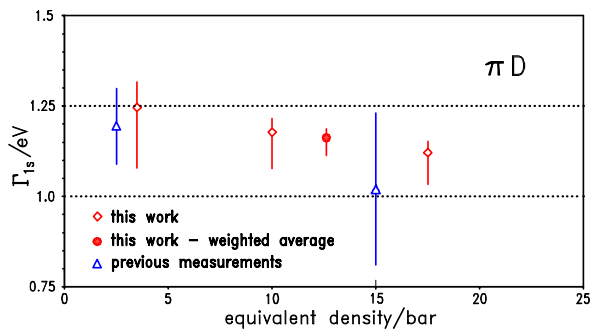}}\vspace{4mm}
\caption{Experimental information on the hadronic broadening in pionic deuterium. 
         Data points of previous measurements are taken from \cite{Cha9597} and \cite{Hau98}.}
\label{figure:width}
\end{figure}

The hadronic broadening is mainly extracted from the sum spectrum of the 10\,bar and 17.5\,bar measurement 
using only a low-energy component for the kinetic energy distribution because no density dependence is 
identified within the experimental accuracy. The combined result 
\begin{eqnarray}
\Gamma_{1s}=\left(1171\,{+\,~23\atop-\,~49}\right)\,\,\mathrm{meV}\label{eq:Ga_final} 
\end{eqnarray}
is obtained by averaging according to the statistical weight. It is in good agreement with the 
earlier measurements, but is a factor of about 3 more precise (Fig.\,\ref{figure:width}).

Evaluating the Deser formula, one obtains for the complex scattering length in leading order (LO) 
\begin{eqnarray}
a^{LO}_{\pi D}=\left(24.8\pm0.3\,-i~6.2 {+0.1 \atop -0.3}\right)\,\cdot 10^{-3}\,m^{-1}_{\pi}\,,
\end{eqnarray}
and the value for the threshold parameter $\alpha$ derived from the $\Im\,a_{\pi D}$ reads
\begin{eqnarray}
\alpha  &=& \left(252{+\,5\atop -11}\right)\,\mu\mathrm{b}\,.
\end{eqnarray}

\begin{figure}[b]
\resizebox{0.45\textwidth}{!}{\includegraphics{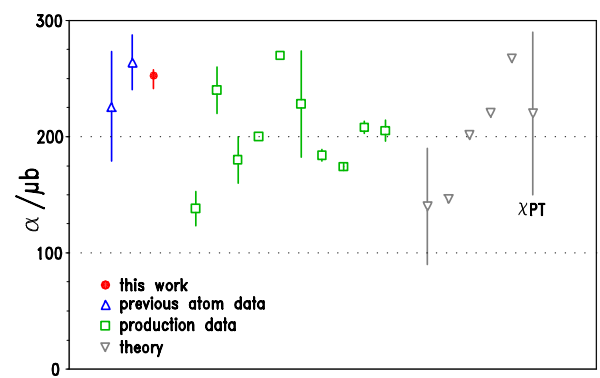}}
\caption{Threshold parameter $\alpha$ for pion production in the reaction $pp\rightarrow d\pi^+$. 
          Input in the Figure (from left to right) are from previous $\pi$D experiments (\cite{Cha9597,Hau98}),
          this experiment, 
          pion production and absorption cross sections 
          (\cite{Cra55,Ros67,Ric70,Hue75,Hue75,Aeb76,Hut90,Rit91,Hei96,Dro98}), 
          and a selection of theoretical approaches (\cite{Ros54,Kol69,Rei69,Afn74,Len06}).}
\label{figure:alpha}
\end{figure}

The parameter $\alpha$ as determined from pion production experiments shows wide fluctuations even when comparing 
recent data\,\cite{Cra55,Ros67,Ric70,Hue75,Aeb76,Hut90,Rit91,Hut91,Hei96,Dro98}. In some cases, only the statistical 
error is given for the cross section of the production experiments, but the fluctuations suggest significant 
systematic uncertainties, which may arise in uncertainties of the normalisation and/or the Coulomb corrections 
(Fig.\,\ref{figure:alpha}). 

\section{Summary}

The $(3p-1s)$ X-ray transition in pionic deuterium has been studied to determine the strong-interaction effects 
with increased precision. An accuracy of 1.3\% was obtained for the shift, which matches with the theoretical 
uncertainty of 1.5\% 
achieved recently\,\cite{Len07,Han09}. The experimental uncertainty is dominated by the error of the calibration 
transition Ga K$\alpha_2$ (1.2\%), which could be decreased by a factor of about 2 by a remeasurement using a 
double flat crystal spectrometer in case a significant theoretical progress requires further improvement. The 
new value for the shift is significantly smaller ($\approx$\,4\%) than the results reported from the two earlier precision 
measurements. About 40\% of the discrepancy is due to the new calculation of the $\pi$D$(3p-1s)$ electromagnetic 
transition energy.

The theoretical understanding of $NN\leftrightarrow NN\pi$ reaction is continuously increasing. Within the 
approach of chiral perturbation theory a study of pion production including the next-to-leading order (NLO) terms 
yields $\alpha^{NLO}=220\,\mu b$\,\cite{Len06,Fil09} in good agreement with the pionic-deuterium result 
(Fig.\,\ref{figure:alpha}). At NLO the theoretical uncertainty is estimated to about 30\%. It is expected 
to decrease below 10\% within a few years from forthcoming NNLO calculations\,\cite{Han09}. Hence, the experimental 
accuracy of 4.2\% for the hadronic broadening in $\pi$D reaches already the expected final 
uncertainty of the theoretical calculations. 

Noteworthy, that at the 10\% level no 
components from high-energy Coulomb transitions could be identified from the fit to the line shape.


\begin{thebibliography}{99}

\bibitem{PSI98}  PSI experiments R-98-01 and R-06-03, www.fz-juelich.de/ikp/exotic-atoms
\bibitem{Got04}  D. Gotta, Prog. Part. Nucl. Phys. {\bf 52} (2004) 133
\bibitem{Got08}  D.\,Gotta et al., Lect. Notes Phys. {\bf 745} (2008) 165
\bibitem{Des54}  S.~Deser, L.~Goldberger, K.~Kaufmann, and W.~Thirring,  Phys. Rev. {\bf 96} (1954) 774
\bibitem{Tru61}  T.~L.~Trueman, Nucl. Phys. {\bf 26} (1961) 57
\bibitem{Ras82}  G.~Rasche and W.~S.~Woolcock, Nucl. Phys. A {\bf 405} (1982) 381
\bibitem{Sig96b} D.~Sigg et al., Nucl. Phys. A {\bf 609} (1996) 310
\bibitem{Lyu00}  V.\,E.\,Lyubovitskij and A.\,Rusetsky, Phys. Lett. B {\bf 494} (2000) 9 
\bibitem{Gas02}  J. Gasser et al., Eur. Phys. J. C {\bf 26} (2002) 13
\bibitem{Zem03}  P.~Zemp, in  {\em Proc. of Chiral Dynamics 2003} , p.~128, Bonn, Germany, September 8--13, 2003, arXiv:hep--ph/0311212v1.
\bibitem{Oad06}  G.\,C.\,Oades, G.\,Rasche, W.S.\,Woolcock, E.\,Matsinos, A.\,Gashi, Nucl. Phys. A {\bf 794} (2007) 73
\bibitem{Gas08}  J.\,Gasser, V.\,E.\,Lyubovitskij, and A.\,Rusetsky, Phys. Rep. {\bf 456} (2008) 167
\bibitem{ThLa80}  A.\,W.\,Thomas and R.\,H.\,Landau, Phys. Rep. B {\bf 58} (1980) 121
\bibitem{Eri88}  T.~E.~O. Ericson and W.~Weise, {\em Pions and Nuclei} (Clarendon, Oxford 1988)
\bibitem{Wei66}  S.~Weinberg, Phys. Rev. Lett. {\bf 17} (1966) 616 
\bibitem{Tom66}  Y.~Tomozawa, Nuovo Cim. A {\bf 46} (1966) 707 
\bibitem{Sig96}  D.~Sigg et al., Phys. Rev. Lett. {\bf 75} (1995) 3245; Nucl. Phys. A {\bf 609} (1996) 269; 
                   erratum A {\bf 617} (1997) 526
\bibitem{Cha9597} D.~Chatellard et al., Phys. Rev. Lett. {\bf 74} (1995) 4157; Nucl. Phys. A {\bf 625} (1997) 855
\bibitem{Hau98}  P.~Hauser et al., Phys. Rev. C {\bf 58} (1998) R1869
\bibitem{Sch01}  H.-Ch.\,Schr\"oder et al., Eur. Phys. J C {\bf 21} (2001) 473

\bibitem{Bar97}  V.~V.~Baru and A.~E.~Kudryavtsev, Phys. of At. Nucl. {\bf 60} (1997) 1475
\bibitem{Bea98}  S.~R.~Beane, V.~Bernard, T.--S.~Lee, and U.--G.~Mei{\ss}ner, Phys. Rev. C {\bf 57}, (1998) 424
\bibitem{Del01}  A.\,Deloff, Phys. Rev. C {\bf 64} (2001) 065205
\bibitem{Eri02}  T.~E.~O.~Ericson, B.~Loiseau and A.~W.~Thomas, Phys. Rev. C {\bf 66} (2002) 014005 
\bibitem{Del03}  A.\,Deloff, {\em Fundamentals in Hadronic Atom Theory}, World Scientific, London (2003)
\bibitem{Bea03}  S.~R.~Beane, V.~Bernard, E.~Epelbaum, U.--G.~Mei{\ss}ner, and D.~R.~Phillips,
                 Nucl. Phys. A {\bf 720} (2003) 399
\bibitem{Bur03}  B.~Burasoy and H.~W.~Grieshammer, Int. J. Mod. Phys. E {\bf 12} (2002) 65
\bibitem{Doe04}  M.~D\"{o}ring, E.~Oset, and M.~J.~Vicente~Vacas, Phys. Rev. C {\bf 70} (2004) 045203
\bibitem{Mei05}  U.-\,G.\,Mei{\ss}ner, U.\,Raha, and, A.\,Rusetsky, Eur. Phys. J. C {\bf 41} (2005) 213 
\bibitem{Mei06}  U.-\,G.\,Mei{\ss}ner, U.\,Raha, and, A.\,Rusetsky, Phys. Lett. B {\bf 639} (2006) 478 
\bibitem{Val06}  M.\,P.\,Valderrama and E.\,R.\,Arriola, arxiv:nucl-th/0605078 (2006) 
\bibitem{Bru51}  K.\,Brueckner, R.\,Serber, and K.\,Watson, Phys. Rev. {\bf 81} (1951) 575
\bibitem{Hig81}  V.~L.~Highland et al., Nucl. Phys. A {\bf 365} (1981) 333
\bibitem{Jos60}  D.~W.~Joseph, Phys. Rev. {\bf 119} (1960) 805
\bibitem{Don77}  R.~MacDonald et al., Phys. Rev. Lett. {\bf 38} (1977) 746
\bibitem{Ros54}  A.\,H.\,Rosenfeld, Phys. Rev. {\bf 96} (1954) 139
\bibitem{Rei69}  A.~Reitan, Nucl. Phys. B {\bf 11} (1969) 170
\bibitem{Mac06}  H.\,Machner and J.\,Niskanen, Nucl.Phys. A {\bf 776} (2006) 172
\bibitem{Bar09}  V.\,Baru, A.\,Rusetski, priv. comm.
\bibitem{Bor80}  E.\,Borie and M.\,Leon, Phys. Rev. A {\bf 21} (1980) 1460
\bibitem{Har90}  F.J.\,Hartmann, {\em Proc. of Physics of Exotic Atoms on Electromagnetic Cascade and Chemistry,  
                 1989, Erice, Italy, 1989} (Plenum Press, New York 1990) p.\,23 and p.\,127, 
                 and references therein
\bibitem{JeMa02} T.\,S.\,Jensen and V.\,E.\,Markushin, Eur. Phys. J. D {\bf 19} (2002) 165; 
                 Eur. Phys. J. D {\bf 21} (2002) 261; Eur. Phys. J. D {\bf 21} (2002) 271
\bibitem{Bet57}  H.~A.~Bethe and E.~E.~Salpeter, {\em Handbuch der Physik} Band {\bf XXXV} (Springer--Verlag, Berlin 1957)
\bibitem{BF78}   L.\,Bracci and G.\,Fiorentini, Nuovo Cim. A {\bf 43} (1978) 9
\bibitem{Czi63}  J.\,B.\,Czirr et al., Phys. Rev. {\bf 130} (1963) 341
\bibitem{Bad01}  A.\,Badertscher et al., Europhys. Lett. {\bf 54} (2001) 313, and references therein
\bibitem{Cov09}  D.~S.~Covita et al., Phys. Rev. Lett. {\bf 102} (2009) 023401 
\bibitem{Covth}  D.\,S.\,Covita, PhD thesis, University of Coimbra (Coimbra 2008)
\bibitem{Taq89}  D.\,Taqqu, AIP Conf. Proc. {\bf 181} (1989) 217
\bibitem{Jon99}  S.\,Jonsell, J.\,Wallenius, and P.\,Froelich, Phys. Rev. A {\bf 59} (1999) 3440 
\bibitem{Lin03}  E.\,Lindroth, J.\,Wallenius, and S.\,Jonsell, Phys. Rev. A {\bf 68} (2003) 032502; 
                 Phys. Rev. A {\bf 69} (2004) 059903(E) 
\bibitem{Kil04}  S.\,Kilic, J.-P.\,Karr, and L.\,Hilico, Phys. Rev. A {\bf 70} (2004) 042506 
\bibitem{Egg65}  J.\,Eggs and K.\,Ulmer, Z. angew. Phys., {\bf 20(2)} (1965) 118 
\bibitem{Zsc82}  G.\,Zschornack, Nucl. Instr. Meth. {\bf 200} (1982) 481 
\bibitem{Nel02}  N.\,Nelms et al., Nucl. Instr. Meth. A {\bf 484} (2002) 419 
\bibitem{Ind06}  P.\,Indelicato et al., Rev. Sci. Instrum. {\bf 77} (2006) 043107 
\bibitem{Ana05}  D.\,F.\,Anagnostopulos et al., Nucl. Instr. Meth. A {\bf 545} (2005) 217 
\bibitem{Tra07}  M.\,Trassinelli et al., J. Phys., Conf. Ser. {\bf 58} (2007) 129 
\bibitem{San98}  M.\,Sanchez del Rio and R.\,J.\,Dejus, Proc. SPIE {\bf 3448} (1998) 246 
\bibitem{Des03}  R.\,Deslattes, E.G.\,Kessler,\,Jr., P.\,Indelicato, L.\,de\,Billy, E.\,Lindroth, and J.\,Anton,
                 Rev. Mod. Phys., vol. {\bf  75}, no. 1, 35 (2003)
\bibitem{Moo}    T.~Mooney, priv. comm.
\bibitem{Str09}  Th.\,Strauch, PhD thesis, Universit\"at zu K\"oln (Cologne 2009)
\bibitem{Str10}  Th.\,Strauch et al. in preparation
\bibitem{Vei73}  WM.J.\,Veigele, At. Data and Nucl. Data Tables {\bf 5} (1973) 51 
\bibitem{Jam75}  F.\,James and M.\,Roos, Comput. Phys. Commun. \textbf{10} (1975) 343 
\bibitem{Hir08}  A.\,Hirtl, PhD thesis, Technische Universit\"at Wien (Vienna 2008)
\bibitem{PP06}   V.\,N.\,Pomerantsev and V.\,P.\,Popov, JETP. Lett. {\bf 83}, 331 (2006); 
                 Phys. Rev. A {\bf 73}, 040501(R) (2006)
\bibitem{JPP07}  T.\,S.\,Jensen, V.\,N.\,Pomerantsev, and V.\,P.\,Popov, arXiv:0712.3010v1 [nucl-th] (2007)
\bibitem{PP07}   V.\,P.\,Popov and V.\,N.\,Pomerantsev, arXiv:0712.3111v1 [nucl-th] (2007)
\bibitem{PDG08}  C.\,Amsler\,et\,al.\,(PDG\,2008),\,Phys.\,Lett.\,B\,{\bf 667}\,(2008)\,1
\bibitem{PI09}   P.\,Indelicato, unpublished
\bibitem{Cra55}  F.~S.~Crawford and M.~L.~Stevenson, Phys. Rev. {\bf 97} (1955) 1305 
\bibitem{Ros67}  C.~M.~Rose, Phys. Rev. {\bf 154} (1967) 1305 
\bibitem{Ric70}  C.~Richard-Serre et al., Nucl. Phys. B {\bf 20} (1970) 413 
\bibitem{Hue75}  J.\,H\"ufner, Phys. Rep. {\bf 21} (1975) 1
\bibitem{Aeb76}  D.~Aebischer et al., Nucl. Phys. B {\bf 108} (1976) 214 
\bibitem{Hut90}  D.~A.~Hutcheon et al., Phys. Rev. Lett. {\bf 64} (1990) 176 
\bibitem{Rit91}  B.~G.~Ritchie et al., Phys. Rev. Lett. {\bf 66} (1991) 568 
\bibitem{Hut91}  D.~A.~Hutcheon et al., Phys. Rev. A {\bf 535} (1991) 618 
\bibitem{Hei96}  P.~Heimberg et al., Phys. Rev. Lett. {\bf 77} (1996) 1012 
\bibitem{Dro98}  M.~Drochner et al., Nucl. Phys. A {\bf 643} (1998) 55 
\bibitem{Kol69}  D.~S.~Koltun and A.~Reitan, Phys. Rev. {\bf 141} (1969) 1413 
\bibitem{Afn74}  I.~R.~Afnan and A.~W.~Thomas, Phys. Rev. C {\bf 10} (1974) 109
\bibitem{Len06}  V.~Lensky et al., Eur. Phys. J. A {\bf 27} (2006) 37 
\bibitem{Len07}  V.\,Lensky et al., Phys. Lett. B {\bf 648} (2007) 46 
\bibitem{Han09}  C.\,Hanhart et al., in preparation
\bibitem{Fil09}  A.\,Filin et al., Phys. Lett. B {\bf 681} (2009) 423

\end{thebibliography}
\end{document}